\documentclass[twocolumn]{svjour3-bb}
\usepackage{amsmath}
\usepackage{amssymb}
\usepackage{cite}
\usepackage{amsfonts}
\usepackage{graphicx}
\usepackage{color}
\newcommand{\bm}[1]{\mbox{\boldmath$#1$}}
\journalname{Acta Rheologica}

\begin{document}
\nonfrenchspacing

\title{
Hydrodynamics of Active Polar Systems in a (Visco)Elastic Background
}
\author{{Harald Pleiner,$^{1,}$\thanks{Harald Pleiner:
pleiner@mpip-mainz.mpg.de}
Daniel Sven{\v s}ek,$^{2}$
and Helmut R. Brand$^{\,1,3}$}}

\date{Received: 3 June 2016 / Revised: 14 July 2016 / Accepted: 25 July 2016\\ 
\copyright\ The Authors 2016. This article is published with open access at Springerlink.com}

\authorrunning{H. Pleiner, 
D. Sven{\v s}ek, and H.R. Brand 
} 
\titlerunning{ 
Active Polar Viscoelastic Systems
}

\institute{
$^1\,$Max Planck Institute for Polymer Research, 55021
  Mainz, Germany\\
$^2\,$Department of Physics, Faculty of Mathematics and Physics,
University of Ljubljana, 
1000 Ljubljana, 
Slovenia\\
 $^3\,$Theoretische Physik III, Universit\"at Bayreuth, 95440 Bayreuth, 
Germany \\
}

\maketitle

\abstract{We derive the full set of macroscopic equations necessary to describe the dynamics of systems with active polar order in a viscoelastic or elastic background. The active polar order is 
 manifested
by a second velocity, whose non-zero modulus is the polar order parameter and whose direction is the polar preferred direction. Viscoelasticity is described by a relaxing strain field allowing for a straightforward change from transient to permanent elasticity.
Relative rotations of the elastic structure with respect to the polar direction are taken into account.
The intricate coupling between active polar order and (transient) elasticity leads to a combined relaxation of the polar order parameter and the strains. The rather involved sound spectrum contains a specific excitation due to a reversible coupling between elasticity and polar order. Effects of chirality are also considered.
}


\section{Introduction  \label{intro}}

The collective dynamics of active systems has attracted increasing attention
of the physics community over the last few years 
(Marchetti et al. 2013).
Here we focus on systems, where the motion of the units 
creates an orientational order (``dynamic order").
Among them are 
schools of fish or flocks of birds 
(Katz et al. 2011; Buhl et al. 2006; Lukeman et al. 2010; Parrish and Edelstein 1999;
Ballerini et al. 2008), 
pattern forming growing bacteria ({\it e.g.} Proteus mirabilis) 
(Cisneros et al. 2007; Loose et al. 2008; Zhang et al. 2010; Fu et al. 2012; 
Watanabe et al. 2002; Moriyama et al. 1995; Yamazaki et al. 2005; Matsushita et al. 1999;
Matsushita et al. 1998), 
biological motors (myosin and actin) (Schaller et al. 2010; Schaller et al. 2011; 
Aditi Simha and Ramasawamy 2002; Hatwalne et al. 2004; Surrey et al. 2001; Nedelec et al. 1997), 
or suitable suspensions of active particles.

These are non-equilibrium systems due to some internal or metabolic chemical reactions that provide the driving forces. They are called active, because 
they are driven individually 
in contrast to conventional driven systems, where an external force is applied globally 
(Cross and Hohenberg 1993). 
If the internal 
 driving force
vanishes, the active systems become passive, with no motion and no orientational order 
(Akhmediev et al. 2013). We will concentrate in the following on the hydrodynamics of the  active state, which can be described rather universally, while the transition to the passive state requires the knowledge (or a model) of the specific driving mechanism (for a recent example cf. (Sven\v sek et al. 2013)). We only consider the polar case, where the active motion and order define a preferred direction that allows for the discrimination between front and rear or backward and forward.

Often the active motion takes place within a passive environment. For a simple fluid background, the hydrodynamic description of active dynamic polar order has been given recently 
(Brand et al. 2013) (for the case of axial order cf. (Brand et al. 2011)).
In this paper
we consider the case of a viscoelastic or a gel-like background that is relevant for growth or motion of many bacteria, and for molecular motor dynamics. We investigate theoretically, in particular, how the elastic degrees of freedom interact with the active ones. 

The active dynamic order should not be mixed up with orientational order in equilibrium systems, like nematic and polar nematic liquid crystals. There, a spontaneous breaking of rotational symmetry leads to a spatial structure that is not related to any external or internal field, nor to any macroscopic motion or movement of the molecules. It is described by a director (or equivalently by a second rank orientational tensor (de Gennes 1975)) or by a polar vector 
(Brand et al. 2006, Brand et al. 2009) for the non-polar and polar case, respectively. These quantities are static in the sense that they do not change under time reversal. This is in contrast to the active dynamic order, where in the polar case the velocity of the active entities is used. We will see that there are characteristic differences in the hydrodynamics 
between systems with static order versus systems with dynamic order.

The macroscopic dynamics of active polar systems in viscoelastic gels contains four key elements. First, there is the active polar ordering that gives rise to an active velocity relative to the passive one. This acts as the order parameter and defines the polar preferred direction (sect. \ref{actpolorder}). Second, in sect. \ref{ATS} we discuss the 2-fluid aspects, like active polar advection and convection, as well as general aspects of the 
implementation of the driving force
(``active stresses"). Third, the viscoelasticity of the passive background is described in sect. \ref{viscoelast} by a relaxing Eulerian strain field. In contrast to the more common approach of an ever more complicated phenomenological constitutive equation for the stress tensor, our description allows for a systematic and thermodynamically direct
coupling of viscoelasticity with dynamic active polar order. Fourth, systems that exhibit
permanent and/or relaxing 
elasticity and have additionally an orientational degree of order, are known to allow for relative rotations, a concept that we discuss for the active polar case in sect. \ref{relrotpol}.

In sect. \ref{actpolgel} we present in detail the macroscopic dynamic equations for achiral active polar systems with a viscoelastic background. We discuss in more detail the combined relaxations of elasticity, relative rotations and active polar order in sect. \ref{combrelax} and the rather involved sound spectrum in sect. \ref{sound}. The paper closes with Conclusions and Perspective 
in sect. \ref{discussion}. Additional features relevant for the chiral case are summarized in 
Appendix A and in Appendix B we compare the results with the case of crystals, for which
the displacement field is a truly hydrodynamic variable.


\section{Key aspects \label{key}}
\subsection{Active polar dynamic order \label{actpolorder}}

The motion of the active entities is described by the macroscopic velocity $\bm{v}_1$. The passive background gives rise to another velocity $\bm{v}_2$. The kinetic energy, therefore, can be written as 
\begin{equation}
\label{kinetic}
\epsilon_{kin}= \frac12 \rho_1 \bm{v}_1^2 + \frac12 \rho_2 \bm{v}_2^2 = \frac{1}{2\rho}  \bm{g}^2+\frac12 \alpha \bm{F}^2
\end{equation}
with $\alpha=\rho_1 \rho_2/\rho = \phi (1-\phi)\rho$,
where $\rho_1=\phi \rho$ and $\rho_2= (1-\phi)\rho$ are the mass 
densities of the active and passive part, respectively,
with $\rho = \rho_1+ \rho_2$ the total mass density and $\phi$ the active mass
fraction.
Accordingly, $\bm{g} = \rho \,\bm{v}$ is the total momentum density defining the mean velocity $\bm{v}$. Of the relative velocity $\bm{F} = \bm{v}_1 - \bm{v}_2$ only its modulus enters eq. (\ref{kinetic}), since $\bm{F}^2=F^2$. The orientation of $\bm{F}$, described by the unit vector $\bm{f} = \bm{F} /F$, is not energetically fixed and describes the spontaneous breaking of rotational symmetry due to the ordered motion of the active entities. This is the appropriate hydrodynamic variable to describe polar dynamic order. Note that it is a true velocity, {\it e.g.} it changes sign under time reversal, opposite to the polarization that is used in the static polar case (Brand et al. 2006).

In contrast to a passive two-fluid description (Pleiner and Harden 2003), with an equilibrium state $\bm{g}= 0 = F$, the active state is characterized by a 
finite $F>0$ due to the active motion. In particular we assume a stationary state $F_s = const. >0$ (and $\bm{g}=0$) as the basic state. It has a finite, non-vanishing active velocity $\bm{F}_s = F_s \bm{f}$. In the co-moving frame, there is $\bm{v} =0$, $\bm{v}_1 = (1-\phi) F_s \bm{f}$, and  $\bm{v}_2 = -\phi F_s \bm{f}$. 

Generally, there can be multiple stationary states that might depend on space, or even non-stationary, time dependent ones, but we will not deal with these cases here. 
We will also not consider a possible transition $F_s \to 0$ to the passive state 
(cf. {\it e.g.} (Vicsek et al. 1995; Sven\v sek et al. 2013)).

The existence of a finite active velocity is due to some internal,
 very often chemical driving force
that {\bf {\it e.g.}} is described by some nonlinear diffusion-reaction equations. This part of the dynamics is system-specific, while we concentrate on the universal aspects. Therefore, we take the finite $F_s$ as given, which is technically achieved by 
an energetic coupling (Brand et al. 2013)
\begin{equation}
\label{activeE}
\epsilon_{act}=- \bm{F}^{act} \cdot \bm{F}=-\alpha F_s F
\end{equation}
to a fictional "active force density" $\bm{F}^{act}= \alpha F_s \bm{f}$. Since $\epsilon_{kin}$ and $\epsilon_{act}$ are the only parts of the total energy density $\epsilon$ that contain $F$, the conjugate is 
\begin{equation}
\label{conjF}
m \equiv \frac{\partial \epsilon}{\partial F} = \alpha (F-F_s)
\end{equation}
and vanishes in the stationary state. A dissipative coupling $\dot F \sim - \xi^\prime m$ describes the relaxation of $F$ to $F_s$.


\subsection{Active transport and stress \label{ATS}}

In an Eulerian description a temporal change of a variable (a field in space and time) can be caused by the advection of that quantity in a velocity field. For vectorial and tensorial fields also convection takes place. As an example we take the concentration dynamics, which is generally of the form (Pleiner and Brand 1996)
\begin{equation}
 \label{phidot}
\dot \phi + v_i\nabla_i \phi + \rho^{-1} \nabla_i j^{\phi}_i = 0
\end{equation}
with the phenomenological current $j^{\phi}_i $.

The advection contribution is reversible, since it shows the same time reversal behavior as $\dot \phi$. In the entropy production it is compensated (together with other transport contributions) by the static isotropic pressure in the diagonal part of the stress tensor 
(Pleiner and Brand 1996). In a usual 1-fluid system, this zero entropy requirement is sufficient to fix this term and no material dependence is possible. 

In a 2-fluid system it is not 
clear
a priori, which velocity has to be used for advection and the zero entropy production requirement is not enough to decide this question. A reasonable procedure 
(Pleiner and Harden 2003) is to use the mean velocity 
for all transport terms, since in that case  
the zero entropy requirement is fulfilled. 
Nevertheless,
the actual advection velocities can be different due to the phenomenological currents. 
In particular, $j^{\phi}_i $ contains a contribution (for the full expression cf. sect. \ref{actpolgel})
\begin{equation}
 \label{phitrans}
j^{\phi,R1}_i =  \gamma_\parallel \rho\,  \phi \,(1-\phi)(F_i - F_s f_i) 
\end{equation}
with the phenomenological reversible transport parameter $\gamma_\parallel$
that leads to the effective, material dependent transport velocity $
\bm{v}_{tr} =
\bm{v} + \gamma_\parallel (1-2\phi) (\bm{F}- F_s \bm{f})$. 
In the stationary state this velocity vanishes in the co-moving frame. In a laboratory frame, 
on the other hand,
where the active velocity is $\bm{v}_1 = F_s \bm{f}$, and the passive one $\bm{v}_2 = 0$, there is a non-vanishing transport velocity $\bm{v}_{tr} = \phi F_s \bm{f}$ rendering the transport term a linear one. Indeed, the linearized active concentration dynamics in the laboratory frame reads
\begin{equation}
 \label{phidotlin}
\dot \phi + \phi F_s f_i \nabla_i \phi + \rho^{-1} \nabla_i j^{\phi,\prime}_i = 0
\end{equation}
with $j^{\phi,\prime}_i $ containing the phenomenological couplings to other degrees of freedom. 

For deviations from the stationary state, $\Delta \bm{v}_{1,2} \equiv \bm{v}_{1,2} - \bm{v}_{1,2}^{stat}$, there is an additional, frame independent contribution to the transport velocity in Eq. (\ref{phidot}) given by $\Delta \bm{v}_{tr} =$
$\gamma^\prime \Delta \bm{v}_1 + \gamma^{\prime\prime} \Delta \bm{v}_2$, with $\gamma^\prime= \phi + \gamma_\parallel (1 - 2\phi)$ and $\gamma^{\prime\prime}= 1- \phi - \gamma_\parallel (1 - 2\phi)$, which is material dependent, but only enters the nonlinear dynamics.

The existence of a finite velocity in the stationary state is a clear indication of the non-equilibrium situation
due to the internal driving force.
The active transport term reflects non-equilibrium, but, like any transport term, cannot drive the system out of equilibrium, since it is reversible with zero entropy production. In the literature 
(Giomi and Marchetti, 2012; Maitra et al. 2014) (and in the Toner-Tu model as presented in 
(Marchetti et al. 2013)) dynamic polar order is often described by a polarization vector, $\bm{p}$, rather than by a velocity-type of variable. The active transport is 
thereby
written as $\bm{v} + w_1 \bm{p}$ where $\bm{v}$ is the passive velocity and $\bm{p}$ is used as the active velocity, with a phenomenological coefficient $w_1$.  Clearly the second part does not transform as a velocity rendering it dissipative
and leading to a non-zero entropy production that drives 
the system. The physical origin of this additional, inadvertent 
driving force
is dubious.

Macroscopic descriptions of active polar or nematic systems show a direct coupling of the stress tensor with the order parameter, often called ``active term". Recently, we showed 
(Brand et al. 2014) that one cannot decide, neither from the symmetry properties of the macroscopic variables involved, nor from the structure of such a cross-coupling, whether the system studied is active or passive. Rather, that depends on whether the variables that give rise to those cross-couplings in the stress tensor are driven 
internally or externally, or not at all. 
Such terms are reversible and come with counter terms that ensure zero entropy production. 
Therefore, they cannot drive the system. On the other hand, if the counter terms are neglected,
the ``active term" leads  to non-zero entropy production and acts as an additional, albeit dubious driving force.


\subsection{Viscoelasticity \label{viscoelast}}

In the rheology literature viscoelasticity is often described by an ad-hoc generalization of Newton's linear relation between viscous stress and velocity gradients, involving nonlinearities, time derivatives and more 
(Oldroyd 1950; Oldroyd 1961; Coleman and Noll 1961; Truesdell and Noll 1965; Giesekus 1966;
Giesekus 1982; Bird et al. 1977; Johnson and Segalman 1977; Johnson and Segalman 1978;
Larson 1988). Such an approach not only immediately becomes unwieldy and intractable when applied to more complicated systems, like the 2-fluid and active and ordered case considered here, it is also rather opaque concerning the compliance with general thermodynamic 
requirements (Pleiner et al. 2005).

Here we will use the genuine hydrodynamic approach to describe elasticity by employing the strain tensor $\varepsilon_{ij}$ as macroscopic variable (Martin et al. 1972; Pleiner and Brand 1991).
In crystals, it
is related to the displacement vector $u_i$, the hydrodynamic variable associated with 
spontaneous breaking of translational symmetry, by $2\varepsilon_{ij} = \nabla_i u_j + \nabla_j u_i$, 
if linearized.
In general (already in crystals exhibiting transport of defects), the strain tensor is 
the appropriate variable.
We will restrict ourselves to linear elasticity, assuming the elastic deformations caused by the active particles to be small.

The linear dynamic elastic equation reads in Eulerian description (Pleiner et al. 2004)
\begin{equation}
 \label{epsdot}
\dot\varepsilon_{ij} - A_{ij}  + v_k\nabla_k \varepsilon_{ij}  +  \varepsilon_{kj} \nabla_i v_k + \varepsilon_{ki} \nabla_j v_k +W_{ij} = 0  
\end{equation}
where $ A_{ij} = (\nabla_i v_j + \nabla_j v_i)/2$ is the ``rate-of-strain" tensor and $W_{ij} $ a phenomenological current. For the transport terms we have kept seemingly nonlinear expressions, since in an active system those terms have linear parts as discussed in the previous section. 

For ordinary, permanent elasticity $W_{ij}$ contains diffusion-like contributions, called vacancy diffusion in crystals, which are usually neglected. For viscoelasticity or transient elasticity, the current $W_{ij}$ contains the relaxation of the strains, 
\begin{equation}
 \label{relax}
W_{ij}^{relax} =  \tau_{ijkl} \Xi_{kl}  
\end{equation}
where $\Xi_{ij} =\partial \epsilon/ \partial \varepsilon_{ij}$ is the elastic stress tensor, the conjugate to the strain tensor, that is obtained as a partial derivative of the the total energy density $\epsilon$. Since $\Xi_{ij}$ contains a contribution $\sim \varepsilon_{ij}$, elastic strains vanish on a finite time scale. Further contributions to $\Xi_{ij}$, describing 
static couplings to other degrees of freedom, are listed in sect. \ref{actpolgel}. 

In addition to relaxation, there are reversible and irreversible contributions to $W_{ij}$ that show in a lucid manner the dynamic couplings of the elastic degree of freedom to the other ones. Using standard hydrodynamic procedures (Pleiner and Brand 1996) for their derivation, one guarantees proper thermodynamic behavior, like positivity of entropy production for dissipative processes and zero entropy production for reversible ones. The full expression for $W_{ij}$ will be presented in sect. \ref{actpolgel}. Here, we only want to discuss the (reversible) seemingly nonlinear contributions
\begin{equation}
\label{epsrevNL}
W_{ij}^{Rnl} = \beta_7 (\varepsilon_{kj} \nabla_i  + \varepsilon_{ki} \nabla_j ) (m f_k ) + \beta_8 m f_k \nabla_k \varepsilon_{ij} 
\end{equation}
with the phenomenological material parameters $\beta_7$ and $\beta_8$. In eq. (\ref{epsdot}) the advection and convective terms are written with the mean velocity $\bm{v}$, but with eq. (\ref{epsrevNL}) these velocities effectively are 
$\bm{v} + \beta_8 \alpha (\bm{F} - \bm{F}_s)$ and $\bm{v} + \beta_7 \alpha (\bm{F} - \bm{F}_s)$, for advection and convection, respectively. As a result they are material dependent. For example, the convection of elastic deformations takes place with the active velocity, $\bm{v}_1$, if $\beta_{7}= 1/\rho_1$, and with the passive one, $\bm{v}_2$, if $\beta_{7}= -1/\rho_2$, but generally it will be something in-between. In the stationary state, which is homogeneous, the convection vanishes. For advection there is a non-vanishing stationary velocity $\phi F_s \bm{f}$ in the laboratory frame. 

The generalization of this hydrodynamic approach to nonlinear viscoelasticity (transient elasticity) using the Eulerian strain tensor has been given in (Temmen et al. 2000; Temmen et al. 2001; Pleiner et al. 2000; Grmela 2002; Pleiner et al. 2004) and its practical usefulness in describing many experiments
has been demonstrated, recently (M\"uller et al. 2016a; M\"uller et al. 2016b). An alternative way of describing visco-elasticity is based on the use of a relaxing orientational order parameter tensor 
(Doi and Edwards 1986; Pleiner et al. 2002) with rather equivalent results 
(Pleiner et al. 2004).


\subsection{Relative Rotations \label{relrotpol}}

Rotations of an elastic structure are related to the displacement vector by $\Omega_{ij} = \tfrac12 (\nabla_i u_j - \nabla_j u_i)$. In isotropic elastomers this variable is irrelevant. If there is an independent preferred direction present, rotations of the preferred direction relative to the elastic medium,
$\tilde{\Omega}_i $,
are additional independent macroscopic degrees of freedom. 
They have been introduced for nematic liquid crystal elastomers by de Gennes 
(de Gennes 1980). This concept
has been included into the macroscopic dynamic 
description of liquid crystalline elastomers (Brand and Pleiner 1994). 
The nonlinear generalization of relative
rotations (Menzel et al. 2007) is vital in describing and explaining
the reorientation behavior of the nematic director under an external mechanical force 
(Menzel et al. 2009a; Menzel et al. 2009b; Urayama 2007; Rogez and Martinoty 2011).

In the present case the preferred direction is ${f_i}$ and
\begin{equation} \label{relrot}
\tilde{\Omega}_i =
\delta f_i - \Omega^{\bot}_i =
\delta f_i - \frac{1}{2}f_j\left(\nabla_iu_j - \nabla_ju_i\right)
\end{equation}
is perpendicular to $f_i$, since $f_i \tilde \Omega_{i}=0$. The dynamic equation can be written as
\begin{equation}
\label{Omegadot}
\dot{\tilde{\Omega}}_i + v_k\nabla_k \tilde{\Omega}_i + Z_i = 0 
\end{equation}
Relative rotations are not truly hydrodynamic variables, since they are neither conserved nor 
related to any spontaneously broken continuous symmetry. 
Therefore, they relax in a finite time and the dissipative part of $Z_i$ contains
\begin{equation}
 \label{Orelax}
Z_{i}^{relax} =  \tau^D  \delta_{ij}^\perp \Sigma_{j}  
\end{equation}
where the relative torque, $\Sigma_{i} =\partial \epsilon/ \partial \tilde \Omega_{i}$, is the conjugate to the relative rotations, and is obtained as a partial derivative of the the total energy density $\epsilon$. The transverse Kronecker symbol, $\delta_{ij}^\perp = \delta_{ij} - f_i f_j$, reflects the transverse nature of $\tilde \Omega$, $\Sigma_i$ and $Z_i$.

The full expression for $Z_i$ will be given below.
In contrast to the nematic case,  $\tilde{\Omega}_i$ has the transformation properties of a velocity and, therefore, shows in $Z_i$ rather different types of couplings to the other variables.


\section{Macroscopic equations \label{actpolgel}}
\subsection{Statics \label{statics}}

We are now in a position to present the full macroscopic dynamic equations for active polar order in a viscoelastic background. In addition to the variables already introduced in sect. \ref{key}, total energy density $\epsilon$, active concentration $\phi$, total mass density $\rho$, total momentum density $\bm{g}$, polar order $F$ and preferred direction $f_i$, elastic strain tensor $\varepsilon_{ij}$, and relative rotations $\tilde \Omega_i$, there is the entropy density $\sigma$. The latter is related to all other variables by the Gibbs relation
\begin{eqnarray}
\label{Gibbsf}
T\, d\sigma  &=& d\epsilon -  \Pi \, d\phi -  \mu\, d\rho - \bm{v} \cdot
d\bm{g}  - m\, dF  - h_i \,d  f_i  \nonumber \\ &-& \Xi_{ij}d\varepsilon_{ij} - \Sigma_{i}d\tilde{\Omega}_i 
\end{eqnarray}
with the thermodynamic conjugates as prefactors, temperature $T$, osmotic pressure $\Pi$, chemical potential $\mu$, the mean velocity $\bm{v}$, the order conjugate $m$, the 'molecular field' $h_i$, the elastic stress tensor $\Xi_{ij}$, and the relative torque $\Sigma_i$. They all follow from the total energy density as partial derivatives. Note that $h_i$ and $\Sigma_i$ have to be transverse ($f_i h_i =0= f_i \Sigma_i$) and $\Xi_{ij}= \Xi_{ji}$ symmetric.

In the absence of any orienting field, the orientation of $f_i$ is energetically not fixed and only $\nabla_j f_i$ enters the Gibbs relation.
It is therefore useful to make this explicit in the Gibbs relation
\begin{eqnarray}
\label{GibbsFi}
d\epsilon &=& T\, d\sigma +  \Pi \, d\phi +  \mu\, d\rho + \bm{v} \cdot
d\bm{g}  + m^\prime dF  + h_i^\prime d  f_i  \nonumber \\ &+& 
\Psi_{ij} d \nabla_j F_i +
\Xi_{ij}d\varepsilon_{ij} +
\Sigma_{i}d\tilde{\Omega}_i 
\end{eqnarray}
with $h_i = h_i^\prime - F \nabla_j \Psi_{ij}$ and $m = m^\prime - f_i\nabla_j \Psi_{ij}$.
\\

To guarantee rotational invariance of the energy, there must be the symmetry
\begin{eqnarray} \label{rotinvar}
&& h_i^\prime \delta f_j + \Psi_{ki} \nabla_j F_k + \Psi_{ik} \nabla_k F_j 
+ 2 \Xi_{ik} \varepsilon_{jk} 
+ \Sigma_i \tilde \Omega_j \nonumber \\ && \hspace{0.5cm}= \{ i \Longleftrightarrow j \} 
\end{eqnarray}

To be meaningful we need a definite expression for the energy density
\begin{eqnarray} \label{E}
\epsilon = \epsilon_{kin} + \epsilon_{act} +  \epsilon_{grad} + \epsilon_{state} +\epsilon_{mix} + 
\epsilon_{elast} + \epsilon_{rot} \quad
\end{eqnarray}
Restricting ourselves to the harmonic approximation, symmetry allows for the expressions
\begin{eqnarray}
 \label{Egrad}
 \epsilon_{grad} &=& \tfrac12 K_{ijkl} (\nabla_j F_i)(\nabla_l F_k) 
 \\
\label{Estate}
 \epsilon_{state} &=& \tfrac12 c_{\rho\rho}(\delta\rho)^2
+ \tfrac12 c_{\sigma\sigma}(\delta\sigma)^2 +
\tfrac12 c_{\phi \phi}(\delta \phi) (\delta \phi) 
\nonumber \\
&+&  c_{\rho \phi}(\delta \rho)(\delta \phi) +
c_{\rho\sigma}(\delta\rho)(\delta\sigma)
+c_{\sigma \phi}(\delta\sigma)(\delta \phi) \quad\quad
 \\ 
 \label{Emix}
\epsilon_{mix} &=& (\sigma^{\sigma}_{ijk}\nabla_k \sigma
+ \sigma^{\rho}_{ijk}\nabla_k \rho
+ \sigma^{\phi}_{ijk}\nabla_k  \phi)(\nabla_iF_j)
 \\
\label{Eelast}
\epsilon_{elast} 
&=&  \tfrac{1}{2} c_{ijkl}\varepsilon_{ij}\varepsilon_{kl}+  (\chi_{ij}^{\sigma} \delta \sigma
+ \chi_{ij}^{\rho} \delta \rho
+ \chi_{ij}^{\phi} \delta \phi) \varepsilon_{ij}
 \\
\label{Erot}
\epsilon_{rot} 
&=&  \tfrac{1}{2}D_1\tilde{\Omega}_i\tilde{\Omega}_i
+ D_2\left(f_j\delta^{\bot}_{ik}
+ f_k\delta^{\bot}_{ij}\right)\tilde{\Omega}_i\varepsilon_{jk} 
\nonumber \\
&+&  D_{ijkl}^F (\nabla_j F_i)(\nabla_l \tilde \Omega_k)
\end{eqnarray}
where $\epsilon_{kin}$ and $ \epsilon_{act}$ have been given in eqs. (\ref{kinetic}) and (\ref{activeE}), respectively. The energy $\epsilon$ is not specific for active dynamic polar order and is the same as for a passive nematic elastomer (replacing $f_i$ by the director $n_i$ everywhere).

The second rank tensors are of the uniaxial symmetric form (with $\delta^{\bot}_{ij}\equiv\delta_{ij} -  f_i  f_j$), {\it e.g.}
\begin{equation} \label{symm2rank}
\chi_{ij} = \chi_{||}f_if_j + \chi_{\bot}\delta^{\bot}_{ij},
\end{equation}
with two generalized susceptibilities, each. 
The third order tensors are odd under time reversal and spatial inversion, {\it e.g.} 
\begin{equation} \label{oddf3rank}
\sigma_{ijk}^\rho =
\sigma^{\rho}_1 f_i f_j f_k + \sigma^{\rho}_2 f_j\delta^{\bot}_{ik}
+ \sigma^{\rho}_3\left( f_i\delta^{\bot}_{jk} +  f_k\delta^{\bot}_{ij}\right),
\end{equation} 
and contain three phenomenological static parameters, each. The three fourth order tensors have different symmetries
\begin{eqnarray} \label{K}
K_{ijkl}
&=& \tfrac{1}{2}
K_1\left(\delta^{\bot}_{ij}\delta^{\bot}_{kl}
+ \delta^{\bot}_{il}\delta^{\bot}_{jk}\right)
+ K_2  f_p\epsilon_{pij} f_q\epsilon_{qkl}
 \\  \nonumber
&+& K_3  f_l f_j\delta^{\bot}_{ik}
+ K_4  f_i f_j f_k f_l
+ K_5  f_i f_k\delta^{\bot}_{jl}
 \\ \nonumber
&+& \tfrac{1}{4} 
 K_6\left(  f_i f_l\delta^{\bot}_{kj}
+  f_j f_k\delta^{\bot}_{il} +  f_i f_j\delta^{\bot}_{kl}
+  f_k f_l\delta^{\bot}_{ij} \right)
\end{eqnarray}
and
\begin{eqnarray} \label{M}
c_{ijkl}
&=&
c_1 \delta^{\bot}_{ij}\delta^{\bot}_{kl} +
c_2 (\delta^{\bot}_{ik}\delta^{\bot}_{jl}+\delta^{\bot}_{il}\delta^{\bot}_{jk})
\nonumber \\
&+&  c_3 f_if_jf_kf_l + c_4(f_if_j\delta^{\bot}_{kl}
+ f_kf_l\delta^{\bot}_{ij}) \nonumber
\\
&+&  c_5 ( f_if_k\delta^{\bot}_{jl} + f_if_l\delta^{\bot}_{jk}
+ f_jf_k\delta^{\bot}_{il} + f_jf_l\delta^{\bot}_{ik} ) \quad
\end{eqnarray}
and
\begin{eqnarray} \label{D}
D_{ijkl}^F
&=& \tfrac{1}{2}
D_1^F \left(\delta^{\bot}_{ij}\delta^{\bot}_{kl}
+ \delta^{\bot}_{il}\delta^{\bot}_{jk}\right)
+ D_2^F  f_p\epsilon_{pij} f_q\epsilon_{qkl}
  \nonumber  \\
&+& D_3^F  f_l f_j\delta^{\bot}_{ik}
+  \tfrac{1}{2} 
 D_4^F f_i \left(  f_l\delta^{\bot}_{kj}
 + f_j\delta^{\bot}_{kl}
 \right)
\end{eqnarray}
containing six generalized Frank coefficients, five elastic moduli, and four mixed-type parameters, respectively.
We do not write down here the explicit expressions for the conjugate quantities, since taking the appropriate derivatives is straightforward. We provide those expressions when necessary ({\it e.g.} in sect. \ref{simplesol}). The order conjugate $m$ is given in eq. (\ref{conjF}).


\subsection{Dynamic equations \label {dynamics}}
 
The variables introduced above follow local dynamic equations, either conservation laws or balance equations. Using the symmetry requirements we get 
\begin{eqnarray}
&& \label{rhodot}
\dot \rho + \nabla_ig_i = 0
\\
&& \label{sigmadot}
\dot \sigma + \nabla_i(\sigma v_i) + \nabla_ij^{\sigma}_i =
\frac{R}{T}
\label{entropynew}
\\
&& \label{gdot}
\dot g_i + \nabla_j ( v_jg_i + \delta_{ij}p
+ \tfrac{1}{2}\Psi_{jk}\nabla_k F_i
+\tfrac{1}{2} \Psi_{ik} \nabla_k F_j \nonumber \\ &&\hspace{1.2cm} - \Xi_{ij}+ \Xi_{kj} \varepsilon_{ik} + \Xi_{ki} \varepsilon_{jk}) +\nabla_j \sigma_{ij} = 0 
\\
&& \label{Fdot}
\dot F + v_i \nabla_i F + X =0 
\\
&& \label{fidoti} 
\dot{  f_i} + v_j \nabla_j  f_i +  f_j \omega_{ij} + Y_i =0 
\end{eqnarray}
with the vorticity tensor $\omega_{ij} = \tfrac12 (\nabla_i v_j - \nabla_j v_i)$.
The dynamic equations for $\phi$, $\varepsilon_{ij}$ and $\tilde \Omega_i$ (with the phenomenological currents $j_i^\phi$, $W_{ij}$, $\Sigma_i$) are given in eqs. (\ref{phidot}), (\ref{epsdot}), and (\ref{Omegadot}). These phenomenological currents,  and the other ones, $j_i^\sigma$, $\Xi_{ij}$, $X$, and $Y_i$, can all be split into a reversible part (superscript $R$) and a dissipative part (superscript $D$) according to their time reversal behavior.

As discussed in sect. \ref{ATS} we have written the transport terms using the mean velocity, while additions are provided by the reversible parts of the phenomenological currents given below. For the origin of the rather lengthy non-phenomenological expression in brackets in eq. (\ref{gdot}) cf.  (Pleiner and Brand 1996; Pleiner et al. 2004).
 The dynamic equation for $\bm{F}$ is $\dot F_i +v_j \nabla_j F_i - F_j \omega_{ij} + X_i = 0$ with $X_i = f_i X + F Y_i$.

The source term in eq. (\ref{sigmadot}), the entropy production $R/T$, is zero for reversible processes and positive for dissipative ones. With the help of the Gibbs relation we get for $R$ as a bilinear function of dissipative currents and generalized forces
\begin{eqnarray} \label{Rpotential}
\int R_\sigma \,dV &=& \int dV\bigl( - j_i^{\sigma, D} \nabla_i T - j_i^{\phi,D} \nabla_i (\Pi/\rho) - \sigma_{ij}^D A_{ij} \nonumber \\   &+& Y_i^D h_i + X^D m + W_{ij}^D \Xi_{ij} + Z_i^D \Sigma_i \bigr) >0 \quad
\end{eqnarray} 
Within linear irreversible thermodynamics, currents and forces are linearly related, which allows us  to use the dissipation function, $R$, 
written as a harmonic function of the forces alone, 
as a potential. In particular, using symmetry arguments again, we find
\begin{eqnarray}
 \label{dissfunct}
2R &=& \kappa_{ij} (\nabla_i T)(\nabla_j T) + D_{ij}
(\nabla_i \Pi^\prime)(\nabla_j \Pi^\prime)
\nonumber \\ &+& \xi_{ijklpq}(\nabla_k\Xi_{ip})(\nabla_l\Xi_{jq}) 
+ \tau^D \delta_{ij}^\perp \Sigma_i\Sigma_j  + b^D \delta_{ij}^\perp \, h_i \, h_j \nonumber
\\ &+& \nu_{ijkl} A_{ij} A_{kl} +  \xi^\prime m^2 
+ 2\mu_{ijk}^T A_{ij} \nabla_k T \nonumber \\ &+& 2 \mu_{ijk}^\Pi A_{ij} \nabla_k \Pi^\prime 
+2\mu_{ijklm}^\Xi A_{ij} \nabla_l \Xi_{km}  
\nonumber  \\  &+& 
 2 D_{ij}^T (\nabla_i \Pi^\prime)(\nabla_j T) + 2\xi^{T}_{ijkl}(\nabla_iT)(\nabla_k\Xi_{jl})
 \nonumber  \\  &+& 2\xi^\Pi_{ijkl}(\nabla_i \Pi^\prime)(\nabla_k\Xi_{jl})+2 \xi \delta^{\bot}_{ij} \Sigma_ih_j  \nonumber \\
 &+& \tau_{ijkl} \Xi_{ij} \Xi_{kl} + 2\Xi_{ij} ( \tau^T_{ij} \delta T + \tau_{ij}^\Pi \delta \Pi^\prime + \tau_{ij}^m m) 
\end{eqnarray}
and the dissipative currents are obtained as partial or functional derivatives, {\it e.g.} 
$X^D = \partial R /\partial m$,  $j_i^{\sigma,D} = - \partial R / \partial \nabla_i T$ or
$W_{ij}^D = {\delta R/\delta \Xi_{ij}} =  {\partial R/\partial \Xi_{ij}}  
-\nabla_k\partial R/\partial\nabla_k\Xi_{ij}$ etc.. We do not write 
down here the explicit expressions for the dissipative currents, and only provide them when necessary.

The second rank tensors have the form of eq. (\ref{symm2rank}), the third rank ones are as 
in eq. (\ref{oddf3rank}), 
the fourth order tensors $\nu_{ijkl}$ and $\tau_{ijkl}$
have the same form as the elasticity tensor eq. (\ref{M}) and the fourth rank tensors
$\xi_{ijkl}^T$ and $\xi_{ijkl}^{\Pi}$ take the form given in eq. (\ref{K}).
The structure of the fifth rank tensor $\mu_{ijklm}^{\Xi}$ and the sixth rank tensor 
$\xi_{ijklpq}$ will be elucidated in Appendix B.

The dissipation function contains the relaxations of the order parameter ($\xi^\prime$), 
of elastic strains ($\tau_{ijkl}$), and of relative rotations ($\tau^D$), already discussed 
in sect. \ref{key}, as well as heat conduction ($\kappa_{ij}$), active concentration diffusion 
($D_{ij}$), viscosity ($\nu_{ijkl}$), rotational viscosity of the polar direction ($b^D$).
There are various interesting cross-couplings, thermo-diffusion ($D_{ij}^T$), 
thermo-strain diffusion ($\xi_{ijkl}^T$), solutal-strain diffusion ($\xi_{ijkl}^\Pi$), 
couplings of flow to temperature ($\mu_{ijk}^T$), to osmotic pressure ($\mu_{ijk}^\Pi$), 
and to strain diffusion ($\mu_{ijklm}^\Xi$), and couplings of strain relaxation to 
temperature ($\tau_{ij}^T$), to concentration  ($\tau_{ij}^\Pi$), and to order parameter 
($\tau_{ij}^m$), and a coupling between relative rotations and polar orientations ($\xi$).
The dissipative flow couplings ($\mu_{ijk}^{T,\Pi}$) 
and ($\mu_{ijklm}^{\Xi}$) 
are specific for a dynamic polar system and are absent in {\it e.g.} passive nematic elastomers. 
The strain diffusion ($\xi_{ijklpq}$) is a sixth rank tensor with 16 independent 
coefficients whose detailed structure will be given in Appendix B.

The reversible parts of the phenomenological currents cannot be derived from a potential. Rather, one writes down all terms that are possible by symmetry and makes sure that appropriate cross-coupling terms cancel each other in the entropy production
\begin{eqnarray} \label{RNull}
0&=& \int dV\bigl( - j_i^{\sigma, R} \nabla_i T - j_i^{\phi,R} \nabla_i (\Pi/\rho) - \sigma_{ij}^R A_{ij} \nonumber \\ && \hspace{0.5cm} + Y_i^R h_i + X^R m + W_{ij}^R \Xi_{ij} + Z_i^R \Sigma_i \bigr) \quad
\end{eqnarray} 
Neglecting some higher order gradient terms we find
\begin{eqnarray}
\label{sigmarev}
j_i^{\sigma,R} &=&  \beta_\parallel f_i m + \beta_\bot \delta_{ij}^\bot h_j + \beta_\bot^\Omega \delta_{ij}^\bot \Sigma_j \ 
\\
\label{phirev} 
j_i^{\phi,R} &=& \gamma_\parallel f_i m + \gamma_\bot \delta_{ij}^\bot h_j + \gamma_\bot^\Omega \delta_{ij}^\bot \Sigma_j  
\\
\label{grev}
\sigma_{ij}^{R} &=&  a_{ij}  m  + \lambda_{ijk}  h_k  + \lambda_{ijk}^\Omega  \Sigma_k  
\\ 
\label{firev} 
Y_i^{R} &=&  \delta_{ij}^\bot (\beta_\bot \nabla_j T +  \gamma_\bot \nabla_j \Pi^\prime + \beta_1 \nabla_j m + \beta^W_{kl}  \nabla_k \Xi_{jl} )  \nonumber \\ &+&  \lambda_{kji}  A_{jk}  + Y_i^{Rnl}
 \\
\label{Frev} 
X^R &=&     \beta_\parallel f_i \nabla_i T +  \gamma_\parallel f_i \nabla_i \Pi^\prime + a_{ij} A_{ij} + \beta_1 \delta_{ij}^\bot \nabla_j h_i \quad\quad \nonumber \\
 &+& \beta_1^\Omega \delta_{ij}^\bot \nabla_i \Sigma_j + \tilde \beta_{kl}^W f_i \nabla_k \Xi_{il} + X^{Rnl}
 \\
\label{Omegarev}
Z_i^R &=& \delta_{ij}^\bot( \beta_\bot^\Omega \nabla_j T + \gamma_\bot^\Omega \nabla_j \Pi^\prime + 
\beta_1^\Omega \nabla_j m) \nonumber \\
&+& \beta^\Omega_{kl} \delta_{ij}^\bot \nabla_k \Xi_{jl} +\lambda_{kji}^\Omega A_{jk} + Z_i^{Rnl}
 \\
\label{epsrev}
W_{ij}^R &=& \tfrac12 (\tilde\beta_{kj}^W  f_i  + \tilde\beta_{ki}^W f_j ) \nabla_k m \nonumber \\ 
&+&  \tfrac12 (\delta_{ik}^\bot \beta^W_{jl} + \delta_{jk}^\bot \beta_{il}^W )  \nabla_l h_k
\nonumber \\
&+& \tfrac12 (\delta_{ik}^\bot \beta_{jl}^{\Omega} + \delta_{jk}^\bot 
\beta_{il}^{\Omega} ) \nabla_l
\Sigma_k +W_{ij}^{Rnl}
\end{eqnarray}
with the second rank tensors of the form of eq. (\ref{symm2rank}), and the generalized flow alignment tensors are
\begin{eqnarray} \label{lambdatens}
\lambda_{ijk} &=& \lambda (\delta_{ik}^\bot f_j + \delta_{jk}^\bot f_i)\\
\lambda_{ijk}^\Omega &=& \lambda^\Omega (\delta_{ik}^\bot f_j + \delta_{jk}^\bot f_i) \label{lambdaOmegatens}
\end{eqnarray}
Relative rotations couple reversibly to all other variables 
($\beta_\perp^\Omega,\gamma_\perp^\Omega,\lambda_{ijk}^\Omega,\beta_1^\Omega,\beta^\Omega_{kl}$), 
except to reorientations of the preferred direction. The latter couple to the elastic degree 
of freedom ($\beta^W_{kl}$) as does the order parameter ($\beta_{1kl}^W$).  
All other terms are already present in systems with dynamic polar order in a simple fluid 
background (Brand et al. 2013). Among them is the so-called active term ($a_{ij}$ in the 
stress term) mentioned in sect. \ref{ATS} and discussed in detail in (Brand et al. 2014).

The nonlinear term $W_{ij}^{Rnl}$ has been used in sect. \ref{viscoelast} to show that it modifies the transport velocities of the elastic strain tensor. The counter terms are $\sim \beta_7$ and $\sim \beta_8$ in eq. (\ref{FrevNL}). Similarly,
 \begin{eqnarray}
\label{firevNL}
Y_i^{Rnl} &=& \beta_3   \delta_{ik}^\bot  f_j  (\nabla_j F_k ) \,m + \beta_4 \delta_{ik}^\bot   f_j \omega_{kj}\, m 
\\
\label{FrevNL}
X^{Rnl} &=& \beta_3 \delta_{ik}^\bot  f_j  (\nabla_j F_k ) h_i  + \beta_4 \delta_{ik}^\bot  f_j \, \omega_{jk} h_i  \nonumber \\
&+& 2 \beta_7 f_k  \nabla_i ( \Xi_{ij} \varepsilon_{kj}) - \beta_8 \Xi_{ij}f_k \nabla_k \varepsilon_{ij} 
\nonumber \\ &-& \beta_5 f_k \Sigma_i \nabla_k \tilde \Omega_i \\
\label{OmegarevNL}
Z_i^{Rnl} &=& \beta_5 m f_k \nabla_k \tilde \Omega_{i}
\end{eqnarray}
and $\beta_5$ adds to the advective velocity of relative rotations in eq. (\ref{Omegadot}), while $\beta_3$ and $\beta_4$ contribute to the advective and convective velocity of $f_i$, eq. (\ref{fidoti}) respectively.


\section{Simple solutions \label{simplesol}}

\subsection{Combined relaxations \label{combrelax}}

The set of macroscopic equations derived above contains three variables that relax to a certain stationary value, even when the deviations are homogeneous in space: Elastic strains $\varepsilon_{ij}$ relax to zero, since elasticity is transient only for a visco-elastic medium, the order parameter of the active polar order $\delta F$ relaxes to its non-zero stationary value $F_s$, and the relative rotations $\tilde \Omega_i$ relax to zero, since they constitute a non-hydrodynamic degree of freedom. 

Actually these three types of variables are coupled and so are their relaxations. The free energy eq. (\ref{Erot}) shows the coupling between relative rotations and elastic strains (via $D_2$), well-known already from nematic elastomers, while the active polar order couples dynamically to elasticity via the very last term ($\sim\tau^m_{ij}$) in the dissipation function eq. (\ref{dissfunct}). In addition, elastic strains are related dynamically to flow, $A_{ij}$, eq. (\ref{epsdot}), and statically to changes of entropy or temperature, density and concentration, eq. (\ref{Eelast}). In the homogeneous case these variables do not have their own dynamics and are simply slaved by the strains. Thus, for homogeneous relaxations we are left with 
 \begin{eqnarray}
\label{relaxgen}
&&\dot \varepsilon_{ij} + \tau_{ijkl} \Xi_{kl} + \tau_{ij}^m (F-F_s) =0  \\
&&\dot{ \tilde{ \Omega_i}} + \tau^D ( D_1 \tilde \Omega_i + D_2 [ f_j \delta_{ik}^\perp + f_k \delta_{ij}^\perp] \varepsilon_{jk}) =0 \\
&&\dot F + \xi^\prime \alpha (F-F_s) + \tau_{ij}^m  \Xi_{ij} =0
\end{eqnarray}
with $\Xi_{ij} = c_{ijkl} \varepsilon_{kl} + D_2 (f_j \delta_{ik}^\perp + f_i \delta_{jk}^\perp) \tilde \Omega_k$.

We notice
that relative rotations only couple to shear strains ($\varepsilon_{ij}$ off-diagonal with respect to $\bm{f}$), while the active polar order couples to compressional strains ($\varepsilon_{ij}$ diagonal with respect to $\bm{f}$). The former case is also present for transient nematic elastomers (nematic polymers), although, to our knowledge, it has not been considered before. It is not related to active order and will be discussed only briefly at the end of this section. We will concentrate here on the coupling to polar active order.
To reduce the algebraic complexity of the equations, and since it is a rather good approximation for polymers, we will assume incompressibility in the strains, $\varepsilon_{ii} = 0$. The remaining two equations are (with $\bm{f}$ along the z-direction)
 \begin{eqnarray}
\label{relaxcomp}
&&\dot \varepsilon_{zz} + \zeta  \varepsilon_{zz} +  \tau_m \alpha (F-F_s) =0  \\
&&\dot F + \xi^\prime \alpha (F-F_s) +  \tau_m c_m \varepsilon_{zz}=0
\end{eqnarray}
with the abbreviations 
$\zeta \equiv  \tau_3 (c_3 - c_4) + 2 \tau_4 (c_4 - c_1 -c_2)$,  $c_m \equiv c_1 +  c_2 +c_3 - 2c_4$, 
and $\tau_m \equiv \tau_{\parallel}^m = -2 \tau_\perp^m$. The latter relation is necessary to maintain incompressibility for all times, $\dot \varepsilon_{ii}=0$. There are appropriate relations for combinations of $\tau$'s and $c$'s, which we will not show here.

The general solution is a coupled 
 sum of two exponential decays
 \begin{eqnarray}
\label{relaxsolution}
\begin{Bmatrix} \varepsilon_{zz} \\ F - F_s \end{Bmatrix} = \begin{Bmatrix} A_1 \\ B_1 \end{Bmatrix}
 \exp(-\lambda_1 t)
+ \begin{Bmatrix} A_2 \\ B_2 \end{Bmatrix}
 \exp(-\lambda_2 t) \quad
\end{eqnarray}
with the inverse relaxation times
\begin{equation} \label{eigenwerte}
\lambda_{1,2} = \frac12 (\zeta + \alpha \xi^\prime) \pm \frac12 \Bigl(  (\zeta - \alpha \xi^\prime)^2 + 4 \alpha c_m \tau_m^2 \Bigr)^{1/2}
\end{equation}
Although the coupling constant $\tau_m$ is bound from above,
$\tau_m^2 < \tau_3 \xi^\prime$,
by the positivity requirement of the dissipation function, it is
not guaranteed that the $\lambda$'s are both positive. This is
achieved by the requirement of the additional relation $\tau_3 = -2 \tau_4$.

The $A$- and $B$-amplitudes are coupled, {\it e.g.} $ \alpha \tau_m B_{1,2}= -(\lambda_{1,2} + \zeta) A_{1,2} $, and any relaxation of one variable triggers the relaxation of the other. Assuming initially $F=F_s$ and a finite deformation $\varepsilon_{zz}(t=0)=A$, the latter not only relaxes to zero with the amplitudes 
 \begin{equation}
\label{initepsA}
A_{1,2} = \pm \,\frac{\zeta- \lambda_{2,1}}{\lambda_1 - \lambda_2} A
\end{equation}
but also induces a transient change in the active polar order 
 \begin{equation}
\label{initepsB}
B_{1}=-  B_2 = \frac{c_m \tau_m }{\lambda_1 - \lambda_2}\, A
\end{equation}
starting with the slope $B(\lambda_2 - \lambda_1)$ and finally relaxing to zero with the longest relaxation time.
Vice versa, an initial order parameter deviation from the stationary value, $F(t=0) - F_s = B$ leads to the appropriate result of an induced transient compression
 \begin{eqnarray}
\label{initF}
A_{1}= - A_2 &=&  \frac{\alpha \tau_m }{\lambda_1 - \lambda_2} \,B \\ \textrm{and} \quad
B_{1,2} &=& \pm \,\frac{\alpha \xi^\prime - \lambda_{2,1}}{\lambda_1 - \lambda_2} B
\end{eqnarray}

The combined relaxation of relative rotations and shear strains, {\it e.g.} $\tilde \Omega_x$ and $\varepsilon_{xz}$, is described by the sum of two exponentials (similar to eq. (\ref{relaxsolution})) with the inverse relaxation times
\begin{equation} \label{eigenwerte2}
\lambda_{3,4} = \tau_5 c_5 + \frac{\tau^D D_1 }{2}  \pm \Bigl(  \bigl[\tau_5 c_5 -  \frac{\tau^D D_1 }{2}  \bigr]^2 + 8 \tau_5 \tau^D D_2^2 \Bigr)^{1/2}
\end{equation}
again indicating that relaxing relative rotations induce shear strains and vice versa.


\subsection{Sound spectrum \label{sound}}

In the preceding section we concentrated on homogeneously
relaxing strains. Here, we will consider the case that either elasticity is permanent (the relaxation times diverge) or that experiments are done on a time scale small compared to the relaxation times. We concentrate on sound-like excitations, {\it i.e.} propagating plane waves ($\sim \exp i (\omega t - \bm{k}\cdot \bm{r})$) characterized by a frequency $\omega$ and a wave vector $\bm{k}$. We only consider effects to order $\omega \sim k$ and neglect dissipative effects. 

In simple liquids the only propagating excitation is ordinary sound, which is based on the reversible coupling between momentum and density dynamics via the density dependence of the pressure (compressibility). It is called longitudinal, since the space and time dependent velocity  is along the wave propagation direction given by $\bm{k}$. 

In nematic and polar nematic systems there is no additional propagating excitation in order $\omega \sim k$, since the nematic order is related to broken 
rotational symmetry 
and the flow alignment coupling ($\lambda_{ijk}$) only leads to $\omega \sim k^2$. With active polar order the situation is completely different (Brand et al. 2013), since the preferred direction is a second velocity. It has been shown that the reversible couplings of entropy, concentration and the stress tensor to the polar order, $\gamma_\parallel$, $\beta_\parallel$, $a_{\parallel}$ in eqs. (\ref{sigmarev}) - (\ref{grev}), lead to a second, generally propagating, longitudinal sound. It is anisotropic, depending on the orientation of $\bm{k}$ relative to $\bm{f}$, and couples to the ordinary sound such that the latter also becomes anisotropic. In addition, it reflects the broken time reversal symmetry due to the non-equilibrium nature of the active polar state (cf. eq. (76) of (Brand et al. 2013)).

In passive elastomers and crystals elasticity allows for additional propagating sound-like excitation. The direct relation between strain dynamics and flow, eq. (\ref{epsdot}), and the elastic stress contribution to the stress tensor, eq. (\ref{gdot}), lead to transverse, shear-elastic sound and an elastic contribution to the longitudinal sound velocity. Only the elastic moduli $c_{ijkl}$ in eq. (\ref{Eelast}) are involved as material parameters.

In the present system we have all these three excitations together, and, of course, they are coupled. 
In addition, there are extra excitations beyond those of the individual subsystems. In particular,
 the direct reversible coupling between elasticity and active polar order, provided by the reactive transport parameter $\beta_1^W$ in eqs. (\ref{Frev}) and (\ref{epsrev}), results in an additional propagating mode with
\begin{equation}
\omega^2_{1,2} = \alpha (\beta_1^{W})^2 ( 4c_5 k_\perp^2 + c_3 k_\parallel^2)
\end{equation}
that also couples to the previously discussed modes. It involves compressional, $\varepsilon_{zz}$, and shear strains $\varepsilon_{xz} + \varepsilon_{yz} $. This mode can be weakly damped 
due to the relaxation of $F$.

There are other reversible cross-coupling terms, like $\beta_1$ in eqs. (\ref{firev}) and (\ref{Frev}), and $\beta^W$ in eqs. (\ref{firev}) and (\ref{epsrev}). They only contribute to the order $\omega \sim k^2$, since in both cases reorientations of the active polar direction are involved and $f_i$ is a 
variable associated with a spontaneously broken continuous symmetry,
of which only gradients enter the free energy. In principle, also the reversible cross-couplings in eq. (\ref{Omegarev}), $\lambda_{ijk}^\Omega$, $\beta_1^\Omega$, and $\beta^\Omega$, of relative rotations with, respectively, flow, eq. (\ref{grev}), polar order, eq. (\ref{Frev}), and elasticity, eq. (\ref{epsrev}), lead to sound-like excitations; in particular to 
$\omega^2 = (D_1/\rho) (\lambda^\Omega)^2 k^2$, $\omega^2 = \alpha D_1 (\beta_1^\Omega)^2 k_\perp^2$ and $\omega^2 = D_1 (\beta^\Omega)^2 ([c_1 + 2c_2] k_\perp^2 + 4 c_5 k_\parallel^2)$. The probably strong relaxation of the relative rotations 
could damp these modes rather quickly.


\section{Conclusions and Perspective \label{discussion}}

We have investigated the influence of a passive transient or permanent network
on systems with a polar dynamic active preferred direction, the latter giving
rise to a second type of fluid and thus
to a second velocity field.
To describe the permanent and/or transient 
networks we use the strain tensor as a macroscopic variable.
We have elucidated the differences to the hydrodynamics of 
crystals for which the displacement vector is a truly hydrodynamic variable.
In bio-inspired systems the presence of a transient or permanent network
is ubiquitous. We just mention the case of bacteria moving on and/or in
an agar gel (Watanabe et al. 2002; Yamazaki et al. 2005).

When writing down the couplings, we pointed out those that are specific 
for a dynamic preferred direction and also commented on those that 
are unchanged with respect to the case of static order 
in the form of a vector or a director.
For the macroscopic description of a system including a network as well as
a second velocity we find intricate cross-coupling terms between the active
polar order and transient elasticity.  
These give rise to a combined relaxation of the polar order parameter and strains 
as well as to an additional propagating mode in the sound spectrum due to 
the reversible coupling between elasticity and polar order.
Relative rotations could, depending on their relaxation times, lead to 
additional sound-like excitations. Or to put it differently, a
sufficiently slow relaxation of relative rotations will lead to propagating modes  
due to the reversible coupling of relative rotations to flow, polar order and elasticity.

In an appendix we have demonstrated that systems that have a polar dynamic preferred direction
as well as a transient or permanent network and are, 
in addition, chiral,  
possess novel reversible cross-coupling terms that lead in particular to reversible
heat and concentration currents. We also note that chirality opens 
up additional dissipation channels.

The present general description of active dynamic polar order with a network 
should be applicable to systems, where two different velocities, for the active and the 
passive part, in addition to a gel, are important. Among those one can think of  
bio-convection of bacteria colonies in a non-Newtonian solvent background or in general of 
bacteria and cells moving in non-Newtonian fluids as they are common
in biological systems on the microscale.

We are aware that the two-fluid description is rather comprehensive. Reductions and simplifications for particular systems are plausible.
Active entities like bacteria and other unicellular organisms, propelled microtubules, insects etc.. are often functioning on or in close proximity of a substrate and in an effective contact with it, which can eventually make the momentum conservation obsolete. In such cases the background velocity can be left out and only the active velocity enters the description 
(Sven\v sek et al. 2013). A possible inclusion of a gel-like background is nevertheless sensible also in these cases. 
Moreover, sometimes even when momentum is conserved, the passive fluid is exempt from consideration, like {\it e.g.} in collective dynamics of birds, fish, flying insects etc.. There the coupling to the background fluid does exist but is normally neglected, typically because the 
``driving" force
is sophisticated and rather autonomous. These systems are driven far from equilibrium, due to the existence of 
sensing and nervous processing/control.
Their properties are also not necessarily due to local interactions
(Pearce et al. 2014).

As a challenge for future work we leave the question to what extent the
approach presented here must be modified to 
macroscopically describe active polar gels, where the strain  formation and 
relaxation is an active process.


\bigskip

\noindent\textit{Acknowledgments:} 
\noindent

H.R.B. and H.P. acknowledge partial support by the 
Deutsche Forschungsgemeinschaft through the Schwerpunktsprogramm
SPP 1681 'Feldgesteuerte Partikel-Matrix-Wechselwirkungen:
Erzeugung, skalen\-\"ubergreifende Modellierung und Anwendung magnetischer 
Hybridmaterialien'. 
D.S. acknowledges the support of the Slovenian Research Agency, Grant No. P1-0099: Physics of soft matter, surfaces, and nanostructures.


\setcounter{equation}{0}
\renewcommand{\theequation}{A.\arabic{equation}}

\section*{Appendix A: Chiral additions \label{chiraladd}}

In chiral systems the inversion symmetry in space is broken. In hydrodynamics the most efficient way of dealing with that is the introduction of a pseudoscalar quantity, $q_0$, that changes sign under inversion. Such a quantity exists either because the units or molecules are chiral by themselves and have a specific handedness, or the order parameter structure is complicated enough to allow for a $q_0$. In the former case the handedness of the system is fixed, while in the latter, generally, both types of handedness are possible and energetically equivalent (ambidextrous chirality) (Brand et al. 2002; Brand and Pleiner 2010). The present case is of the first type and the handedness can originate from the active as well as the passive part. 

For the energy density we have chiral contributions that add to the sum of energies, eq. (\ref{E}),
\begin{eqnarray}
\epsilon_{chir}
&=& 
q_0  \tilde K_2   {\bf f} \cdot ( \nabla \times {\bf F})\nonumber \\
&-&  q_0 
 {\bf f} \cdot  (\nabla \times {\bf F})
( \tau_\phi \delta \phi + \tau_{\sigma} \delta\sigma +
\tau_{\rho} \delta\rho + \tau_\varepsilon \varepsilon_{ii}) \nonumber 
 \\ &+& q_0 \tau^\Omega   f_jf_m \epsilon_{imk}  \tilde \Omega_i \nabla_j F_k
\label{energychiral}
\end{eqnarray}
describing first linear twist, responsible for the helical structure of the preferred direction to be the energetic minimum state (de Gennes 1975). Second are the static Lehmann-type couplings 
(Brand and Pleiner 1988; Brand and Pleiner 2001) to scalar variables and to elastic compressional deformations. The final term is operational only for distorted helical structures.
This last contribution is a higher order gradient term.
It also exists in the usual active polar systems without viscoelastic effects 
(Brand et al. 2013).

The energy $\epsilon_{chir}$ is the same as for a passive chiral nematic elastomer (replacing $f_i$ by the director $n_i$ everywhere).

In the dissipation the chiral additions to eq. (\ref{dissfunct}) are, neglecting some higher order gradient terms,
\begin{eqnarray}
\label{dissipation-chiral}
R_{chir}  &=& q_0 f_i \epsilon_{ijk} (\psi_{j}^{(1)}  \Sigma_k + \psi_{j}^{(2)} h_k) + q_0 \psi^{(3)}_{ij}   \Xi_{ij} \nonumber \\
&+&
q_0 \kappa_{ijk}^\parallel (\zeta_1^A \Sigma_i + \zeta_2^A h_i) A_{jk} \nonumber \\
&+& q_0 (\zeta_\parallel^A \kappa_{jklm}^\parallel +
\zeta_\bot^A  \kappa_{jklm}^\bot )A_{jk} \Xi_{lm} 
\end{eqnarray}
where
\begin{eqnarray}
\label{psi12}
\psi_{j}^{(1,2)} &\equiv& \psi_{1,2}^T \nabla_j T + \psi_{1,2}^\Pi \nabla_j \Pi^\prime  + \psi_{1,2}^m \nabla_j m \quad\quad 
\\
\label{psi3}
\psi^{(3)}_{ij} &\equiv& \sum\nolimits_B (\psi_{3\parallel}^B \kappa_{kij}^\parallel + \psi_{3\bot}^B \kappa_{kij}^\bot) \nabla_k B 
\end{eqnarray}
for $B= T,\Pi^\prime,m$ with
\begin{eqnarray}
\label{e3par}
\kappa_{ijk}^\parallel &=& \tfrac12 (f_l f_k \epsilon_{ijl} + f_l f_j \epsilon_{ikl}) 
\\
\label{e3bot}
\kappa_{ijk}^\bot &=& \tfrac12 (\delta_{lk}^\bot \epsilon_{ijl} + \delta_{lj}^\bot \epsilon_{ikl})
\\
\label{e5par}
\kappa_{jklm}^\parallel &=& \tfrac14 f_i (f_k f_m \epsilon_{ijl} + f_l f_j \epsilon_{ikm} \nonumber \\
&& \hspace{0.4cm} + f_j f_m \epsilon_{ikl} + f_l f_k \epsilon_{ijm}) 
\\
\label{e5bot}
\kappa_{jklm}^\bot &=& \tfrac14 f_i (\delta_{km}^\bot \epsilon_{ijl} + \delta_{jl}^\bot \epsilon_{ikm} + \delta_{jm}^\bot \epsilon_{ikl} + \delta_{kl}^\bot \epsilon_{ijm}) \quad\quad
\end{eqnarray}

The first line contains generalized dynamic Lehmann contributions 
(Brand and Pleiner 1988; Brand and Pleiner 2001; Sven\v sek et al. 2008) that relate gradients of temperature, osmotic pressure and order deviations with relative rotations ($\psi_i^{(1)}$), reorientations of the preferred direction ($\psi_i^{(2)}$), and relaxing strains ($\psi^{(3)}_{ij}$). Such terms also exist in passive systems. The chiral couplings of flow to relative rotations ($\sim\zeta^A_1$), to reorientations of the preferred direction ($\sim\zeta^A_2$), and to relaxing strains ($\sim\zeta^A_{\parallel,\perp}$) are specific for a dynamic order and do not exist in passive systems.

For the chiral additions to the reversible currents, we first have
\begin{eqnarray}
\label{sigmarevchir}
j_i^{\sigma Rc} &=& q_0 g^{(1)}_{ijk} A_{jk} + q_0 g^{(2)}_{ijklm} f_k \nabla_l \Xi_{jm}  
\\
\label{phirevchir}
j^{\Phi Rc}_i &=& q_0 g^{(3)}_{ijk} A_{jk} + q_0 g^{(4)}_{ijklm} f_k \nabla_l \Xi_{jm}  
\\
\label{grevchir}
\sigma_{ij}^{Rc} &=& -q_0 g^{(1)}_{kij} \nabla_k T - q_0 g^{(3)}_{kij} \nabla_k \Pi^\prime  
\\
\label{epsrevchir}
W_{ij}^{Rc} &=& 
\nonumber  q_0  \nabla_l[  g^{(2)}_{miklj} 
f_k \nabla_m T] \\ 
&+& 
\nonumber  q_0 \nabla_l[  g^{(4)}_{miklj} 
f_k \nabla_m \Pi^{\prime}] \\ 
&+& 
q_0 \kappa_{kij}^\parallel ( g^{(5)} \Sigma_k +  g^{(6)}  h_k )
\\
\label{Omegarevchir}
Z_i^{Rc} &=& -q_0 g^{(5)} \kappa_{ijk}^\parallel \Xi_{jk}  
\\
\label{firevchir}
Y_i^{Rc} &=& -q_0 g^{(6)} \kappa_{ijk}^\parallel  \Xi_{jk}  
\end{eqnarray}
where for $n=1,3$
\begin{equation}
g_{ijk}^{(n)} = g_n^\parallel \kappa_{ijk}^\parallel + 
g_n^\bot \kappa_{ijk}^\bot 
\end{equation}
and for $n=2,4$
\begin{eqnarray}
g_{ijklm}^{(n)} &=& g^{n1} (\varepsilon_{ijk} \delta_{lm}^{\bot} 
+  \varepsilon_{imk} \delta_{lj}^{\bot}) \nonumber \\
&+& g^{n2} (\varepsilon_{jkl} \delta_{im}^{\bot} 
+  \varepsilon_{mkl} \delta_{ij}^{\bot}) \nonumber \\
 &+& g^{n3} \varepsilon_{ikl} \delta_{jm}^{\bot} 
 + g^{n4} \varepsilon_{ikl} f_j f_m 
\nonumber \\ 
&+& g^{n5} (\varepsilon_{ijk} f_l f_m 
+  \varepsilon_{imk} f_l f_j) \nonumber \\
&+& g^{n6} (\varepsilon_{ijl} f_k f_m 
+  \varepsilon_{iml} f_k f_j) \nonumber \\
&+& g^{n7} (\varepsilon_{jkl} f_i f_m 
+  \varepsilon_{mkl} f_i f_j) 
\end{eqnarray}
describing mutual couplings of elasticity with temperature, osmotic pressure, relative rotations, 
and reorientations of the preferred direction ($g^{(2)}_{ijklm}, g^{(4)}_{ijklm},g^{(5)},g^{(6)}$, respectively), 
while the flow couplings ($g^{(1,3)}_{ijk}$) are known already from chiral active polar systems with a simple fluid background (Brand et al. 2013). 

In the case of a crystal (cf. Appendix B) these
chiral additions to the reversible currents take the simpler form
\begin{eqnarray}
\label{sigmarevchirxtal}
j_i^{\sigma Rc} &=& \dots  + q_0 g_2 \epsilon_{ijk} f_k \nabla_l \Xi_{jl}  
\\
\label{phirevchirxtal}
j^{\Phi Rc}_i &=& \dots + q_0 g_{4} \epsilon_{ijk} f_k \nabla_l \Xi_{jl}  
\\
\label{epsrevchirxtal}
W_{ij}^{Rc} &=& 
\nonumber 
\dots + \frac{1}{2} q_0 g_{2}  (\epsilon_{ikl} \nabla_j[ f_l \nabla_k T] 
+  \epsilon_{jkl} \nabla_i [f_l \nabla_k T]) \\
&+& 
 \frac{1}{2} q_0 g_{4} (\epsilon_{ikl} \nabla_j[ f_l \nabla_k \Pi^{\prime}] 
+  \epsilon_{jkl} \nabla_i [f_l \nabla_k \Pi^{\prime}]) \quad\quad\quad
\end{eqnarray}
Thus we see that the seven coefficients involved in the reversible coupling
of temperature and concentration gradients to elasticity diffusion
for gels, rubbers and disordered solids reduce to only one independent
contribution in the case of crystals.

Second, there are self-coupling terms, a new one involving relative rotations ($d_4$), in addition to that of flow ($d_3$) and of reorientations of the preferred direction $(d_{5})$, known before. The appropriate self-couplings of the thermal ($d_1$) and soluble ($d_2$) degree of freedom are surface terms
\begin{eqnarray}
j_i^{\sigma Rcn} &=& q_0 d_1 \epsilon_{ijk} f_k \nabla_j T 
\label{sigmarevchirn}
 \\
j^{\Phi Rcn}_i &=& q_0 d_{2} \epsilon_{ijk} f_k \nabla_j \Pi^\prime 
\label{phirevchirn}
\\
\sigma_{ij}^{Rcn} &=& q_0 (d_{3}^\parallel \kappa_{ijlm}^\parallel + 
d_{3}^\bot \kappa_{ijlm}^\bot)  A_{lm} 
\label{grevchirn}
\\
\label{epsrevchirrn}
W_{ij}^{Rc} &=&  q_0 (d_{6}^\parallel \kappa_{ijlm}^\parallel + 
d_{6}^\bot \kappa_{ijlm}^\bot)  \Xi_{lm} 
\\
Z_i^{Rcn} &=& q_0 d_4 \epsilon_{ijk} f_k \Sigma_j 
\label{Omegarevchirn}
\\
Y_i^{Rcn} &=& q_0 d_5 \epsilon_{ijk} f_k h_j 
\label{firevchirn}
\end{eqnarray}
These terms do not have a counterpart, but are nilpotent in the entropy 
production. They are either associated with local distortions of the helical structure or vanish when averaged over many pitch lengths.

The reversible contributions to the stress tensor in eq. (\ref{grevchirn})
have been found first
for the superfluid $A$ phase of $^3$He (Liu 1976). They also arise for
$^3$He-$A_1$ (Pleiner and Graham 1976; Liu 1979), $^3$He-$A$ in high magnetic fields
(Brand and Pleiner 1981) and in $^3$He-$B$ in high magnetic fields (Pleiner and Brand 1983).
They always contain hydrodynamic contributions from the
memory matrix (Brand et al. 1979; Brand and Pleiner 1982).
These reversible contributions to the stress tensor
also arise for uniaxial magnetic gels (Bohlius et al. 2004)
as well as for ferrocholesterics in a magnetic fields (Jarkova et al. 2001).
In the latter case one obtains eight independent coefficients
due to the lower symmetry of the system (Jarkova et al. 2001).
Analyzing the structure of these terms in detail one arrives at the
conclusion that for these contributions to arise
one needs either a preferred direction, which is
odd under time reversal and even under parity
(Liu 1976; Pleiner and Graham 1976; Liu 1979; Brand and Pleiner 1981; 
Pleiner and Brand 1983; Brand et al. 1979; Brand and Pleiner 1982; Bohlius et al. 2004)
or a preferred direction, which is odd under time reversal as
well as odd under parity, but has in addition a pseudoscalar quantity,
as is the case studied here and in ref. (Brand et al. 2013).
In all cases there are reversible contributions from the memory matrix.


\setcounter{equation}{0}
\renewcommand{\theequation}{B.\arabic{equation}}

\section*{Appendix B: Dynamic processes associated with the strain
tensor in gels and rubbers versus
dynamic effects associated with the displacement field  in crystals \label{straindiffusion}}

In the bulk part of this paper we have taken the six components of the strain
tensor $\varepsilon_{ij}$ as independent macroscopic variables.
Such an approach is appropriate for gels, rubbers, transient networks and also
for solids with defects.

The purpose of this appendix is to show how these results reduce to those
of a crystal, which is described by using the displacement field $\vec u$ as strictly
hydrodynamic variables (Martin et al. 1972). Since we deal here with linear elasticity
the relation between the strain tensor and the displacement field simply reads 
$\varepsilon_{ij} = \frac{1}{2} (\nabla_i u_j + \nabla_j u_i)$ and certain 
compatibility conditions guaranteeing the existence of a displacement field are
satisfied. For a detailed discussion of the relation between the displacement field and the
nonlinear strain tensor in the Eulerian picture 
we refer to the discussion in the Appendix of ref. (Pleiner et al. 2000). 

For the static aspects the discussion from 
sect. \ref{statics} remains 
unchanged when making the replacement 
$\varepsilon_{ij} = \frac{1}{2} (\nabla_i u_j + \nabla_j u_i)$ given above.

For the dynamic aspects the situation is different.
While the use of the thermodynamic conjugate $\Xi_{ij}$ can be taken over unchanged 
for reversible and irreversible currents as well as for the 
dissipation function, this is different for its gradients for which we had
in the bulk part $\nabla_i \Xi_{jk}$, without any restriction for the three indices
$i, j$ and $k$.
Since the displacement field is a
truly hydrodynamic variable in a crystal, only the expression 
$\nabla_k \Xi_{jk}$ enters the hydrodynamic description (Martin et al. 1972).
In addition there is no term $\sim \Xi_{ij} $ without gradients in the
dissipation function, since the displacement field is truly hydrodynamic.
Therefore the last line of eq. (\ref{dissfunct}) vanishes completely.
Thus we have for the simplified dissipation function in a crystal for the parts
associated with the displacement field
\begin{eqnarray}
 \label{dissfunctxtal}
2R &&= \dots + 
\xi_{ij}(\nabla_k\Xi_{ik})(\nabla_l\Xi_{jl}) 
 +2\mu_{ijk}^\Xi A_{ij} \nabla_l \Xi_{kl}  
\nonumber  \\ & &+ 
 2\xi^{T}_{ij}(\nabla_iT)(\nabla_k\Xi_{jk})
+ 2\xi^\Pi_{ij}(\nabla_i \Pi^\prime)(\nabla_k\Xi_{jk}) \quad
\end{eqnarray}
The second rank tensors have the form of eq. (\ref{symm2rank}) and carry 2 coefficients each, while the third rank tensor is of the form eq. (\ref{oddf3rank}) with 3 coefficients.

Using the strain tensor instead, there are material tensors of rank four, five, and six involved in eq. (\ref{dissfunct}), carrying much more coefficients. 
In particular we find for 
\begin{equation}
R_5 = \frac12 \mu_{ijklm}^\Xi A_{ij} \nabla_k \Xi_{lm} \\
\end{equation}
twelve coefficients using the symmetries $i \leftrightarrow j$ and $l \leftrightarrow m$, 
\begin{eqnarray} \label{5er}
&& \hspace{-0.5cm}  \mu_{ijklm}^\Xi = \mu_1^\Xi f_k \delta^\perp_{ij} \delta^\perp_{lm}  
+ \mu_2^\Xi (\delta^\perp_{il} \delta^\perp_{jm} + \delta^\perp_{jl} \delta^\perp_{im} ) f_k \cr 
&+& \mu_3^\Xi \delta^\perp_{lm} (\delta^\perp_{kj} f_i + \delta^\perp_{ki} f_j) + \mu_4^\Xi \delta^\perp_{ij} (\delta^\perp_{kl} f_m + \delta^\perp_{km} f_l) \cr
&+& \mu_5^\Xi [(\delta^\perp_{kl}\delta^\perp_{jm} + \delta^\perp_{km} \delta^\perp_{jl})f_i + (\delta^\perp_{kl}\delta^\perp_{im} + \delta^\perp_{km} \delta^\perp_{il})f_j ] \cr
&+& \mu_6^\Xi  [(\delta^\perp_{ik} \delta^\perp_{jl} + \delta^\perp_{jk} \delta^\perp_{il} ) f_m +  (\delta^\perp_{ik} \delta^\perp_{jm} + \delta^\perp_{jk} \delta^\perp_{im} ) f_l ] \cr
&+& \mu_7^\Xi \delta^\perp_{ij} f_k f_l f_m + \mu_8^\Xi \delta^\perp_{lm} f_i f_j f_k +   \mu_9^\Xi f_i f_j f_k f_l f_m \cr
&+& \mu_{10}^\Xi (\delta^\perp_{ik} f_j + \delta^\perp_{jk} f_i) f_l f_m + \mu_{11}^\Xi (\delta^\perp_{kl} f_m + \delta^\perp_{km} f_l)f_i f_j \cr
&+& \mu_{12}^\Xi (\delta^\perp_{il} f_m f_j+ \delta^\perp_{im} f_l  f_j + \delta^\perp_{jl} f_m f_i+ \delta^\perp_{jm} f_l  f_i ) f_k \quad\quad
\end{eqnarray}
and for 
\begin{equation}
R_6 = \frac12 \xi_{ijklpq} (\nabla_k \Xi_{ip})(\nabla_l \Xi_{jq})
\end{equation}
sixteen coefficients 
using the symmetries $k \leftrightarrow l$, $i \leftrightarrow p$, and $j \leftrightarrow q$, 
as well as $i \leftrightarrow j \wedge p \leftrightarrow q$
\begin{eqnarray}  \label{6er}
&&   \hspace{-0.5cm} \xi_{ijklpq} = 
\xi_1 \delta^\perp_{ip} \delta^\perp_{jq} \delta^\perp_{lk}
\cr 
&+&  \xi_2 ( \delta^\perp_{ij} \delta^\perp_{pq} + \delta^\perp _{iq} \delta^\perp_{jp} ) \delta^\perp_{lk}
\cr
&+& \xi_3 ( [\delta^\perp_{ik} \delta^\perp_{lp} + \delta^\perp _{il} \delta^\perp_{pk} ] 
\delta^\perp_{jq} +[ \delta^\perp_{jk} \delta^\perp_{ql} + \delta^\perp _{jl} \delta^\perp_{qk} ] 
\delta^\perp_{ip})
\cr
&+& \xi_4 ( \delta^\perp_{ij} \delta^\perp_{lp} \delta^\perp_{qk} + \delta^\perp_{pj} 
\delta^\perp_{li} \delta^\perp_{qk} +\delta^\perp_{iq} \delta^\perp_{lp} \delta^\perp_{jk} 
+ \delta^\perp_{pq} \delta^\perp_{li} \delta^\perp_{jk} 
\cr
   &&\hspace{0.2cm} + \delta^\perp_{ij} \delta^\perp_{kp} \delta^\perp_{ql} 
+ \delta^\perp_{pj} \delta^\perp_{ki} \delta^\perp_{ql} 
+\delta^\perp_{iq} \delta^\perp_{kp} \delta^\perp_{jl} 
+ \delta^\perp_{pq} \delta^\perp_{ki} \delta^\perp_{jl} )
\cr
&+& \xi_5 ( f_i f_p \delta^\perp_{jq} \delta^\perp_{lk} 
+ \delta^\perp_{ip} f_j f_q \delta^\perp_{lk})
 + \xi_6 \delta^\perp_{ip} \delta^\perp_{jq} f_l f_k 
\cr
&+&   \xi_7 ( \delta^\perp_{ij} \delta^\perp_{pq} + \delta^\perp _{iq} \delta^\perp_{jp} ) f_l f_k
\cr
&+& \xi_8 ( f_i f_j \delta^\perp_{pq} + f_i f_q \delta^\perp_{jp} 
+ \delta^\perp_{ij}  f_p f_q + \delta^\perp _{iq} f_j f_p) \delta^\perp_{lk} 
\cr
&+& \xi_9 ( [f_i f_k \delta^\perp_{lp} + f_i f_l \delta^\perp_{pk} 
+\delta^\perp_{ik} f_l f_p + \delta^\perp _{il} f_p f_k ] \delta^\perp_{jq}
\cr
&& \hspace{0.23cm} +[ f_j f_k \delta^\perp_{ql}     
+ f_j f_l \delta^\perp_{qk} +\delta^\perp_{jk} f_q f_l + \delta^\perp _{jl} f_q f_k ] \delta^\perp_{ip})
      \cr
&+& \xi_{10}  ( [\delta^\perp_{ik} \delta^\perp_{lp} + \delta^\perp _{il} \delta^\perp_{pk} ] f_j f_q 
+[ \delta^\perp_{jk} \delta^\perp_{ql} + \delta^\perp _{jl} \delta^\perp_{qk} ] f_i f_p ) 
\cr
&+& \xi_{11} ( f_i f_j \delta^\perp_{lp} \delta^\perp_{qk} +\delta^\perp_{ij} f_l f_p \delta^\perp_{qk} 
+\delta^\perp_{ij} \delta^\perp_{lp} f_q f_k
\cr 
&&  \hspace{0.28cm} + f_p f_j \delta^\perp_{li} \delta^\perp_{qk} 
+ \delta^\perp_{pj} f_i f_l \delta^\perp_{qk}+ \delta^\perp_{pj} \delta^\perp_{li} f_k f_q
\cr 
&&   \hspace{0.28cm}+f_i f_q \delta^\perp_{lp} \delta^\perp_{jk} 
+\delta^\perp_{iq} f_l f_p \delta^\perp_{jk}  +\delta^\perp_{iq} \delta^\perp_{lp} f_j f_k 
\cr 
&&   \hspace{0.28cm}+ f_p f_q \delta^\perp_{li} \delta^\perp_{jk}
+ \delta^\perp_{pq} f_i f_l \delta^\perp_{jk} + \delta^\perp_{pq} \delta^\perp_{li} f_j f_k
\cr
 &&  \hspace{0.28cm}+ f_i f_j \delta^\perp_{kp} \delta^\perp_{ql} 
+ \delta^\perp_{ij} f_k f_p \delta^\perp_{ql} + \delta^\perp_{ij} \delta^\perp_{kp} f_l f_q 
\cr 
&&  \hspace{0.28cm} + f_p f_j \delta^\perp_{ki} \delta^\perp_{ql}  
+ \delta^\perp_{pj} f_i f_k \delta^\perp_{ql}  + \delta^\perp_{pj} \delta^\perp_{ki} f_l f_q
 \cr 
&&   \hspace{0.28cm}+f_i f_q  \delta^\perp_{kp} \delta^\perp_{jl} 
+\delta^\perp_{iq} f_k f_p \delta^\perp_{jl} +\delta^\perp_{iq} \delta^\perp_{kp} f_j f_l
    \cr 
&&   \hspace{0.28cm}+ f_p f_q \delta^\perp_{ki} \delta^\perp_{jl} 
+ \delta^\perp_{pq} f_k f_i \delta^\perp_{jl}+ \delta^\perp_{pq} \delta^\perp_{ki} f_j f_l)
\cr
&+&\xi_{12} \delta^\perp_{lk} f_i f_p f_j f_q 
\cr 
&+& \xi_{13}(\delta^\perp_{jq}  f_i f_p f_l f_k + \delta^\perp_{ip} f_j f_q f_l f_k )
\cr
&+& \xi_{14} (\delta^\perp_{pq}  f_i f_j +\delta^\perp_{jp}   f_i f_q 
+ \delta^\perp_{ij}  f_p f_q + \delta^\perp _{iq} f_j f_p) f_l f_k
\cr
&+&\xi_{15} ([\delta^\perp_{ki} f_p  f_l +  \delta^\perp_{kp}  f_i  f_l  
+ \delta^\perp_{li}  f_p f_k  + \delta^\perp_{lp} f_i  f_k ] f_j f_q
\cr 
&&   \hspace{0.28cm}
+[ \delta^\perp_{ql}  f_j f_k  + \delta^\perp_{jl} f_q f_k   + \delta^\perp_{jk}  f_l  f_q  
+ \delta^\perp_{qk}  f_j f_l ]f_i f_p  )
\cr
&+& \xi_{16} f_i f_j f_k f_l f_p f_q
\end{eqnarray}

For the reversible currents containing gradients of the elastic stress tensor
we obtain for crystals the slightly simplified expressions
\begin{eqnarray}
\label{firevxtal} 
Y_i^{R} &=&  \dots + \delta_{ij}^\bot  \beta^W \nabla_k \Xi_{jk}   \nonumber 
 \\
\label{Frevxtal} 
X^R &=&  \dots 
  + \beta_1^W f_i \nabla_k \Xi_{ik} 
 \\
\label{Omegarevxtal}
Z_i^R &=& \dots
 +  \beta^\Omega \delta_{ij}^\bot \nabla_k \Xi_{jk}
 \\
\label{epsrevxtal}
W_{ij}^R &=& \tfrac12 \beta_1^W ( f_i \nabla_j m + f_j \nabla_i m) \nonumber \\ 
&+& \tfrac12 (\delta_{ik}^\bot \nabla_j + \delta_{jk}^\bot \nabla_i)
(\beta^\Omega \Sigma_k - \beta^W h_k) 
\end{eqnarray}
Thus instead of obtaining second rank tensors for the case of the strain field 
as a variable, we get a scalar coefficient each for the case of
a displacement field as a truly hydrodynamic variable.\\[0.2cm]


\end{document}